\documentclass[sigconf] {acmart}

\AtBeginDocument{%
  \providecommand\BibTeX{{%
    \normalfont B\kern-0.5em{\scshape i\kern-0.25em b}\kern-0.8em\TeX}}}

\copyrightyear{2024}
\acmYear{2024}
\setcopyright{rightsretained}\acmConference[EICS Companion '24]{Companion of the16th ACM SIGCHI Symposium on Engineering Interactive Computing Systems}{June 24--28, 2024}{Cagliari, Italy}
\acmBooktitle{Companion of the16th ACM SIGCHI Symposium on Engineering Interactive Computing Systems (EICS Companion '24), June 24--28, 2024, Cagliari, Italy}
\usepackage[ruled]{algorithm2e}

\usepackage{xspace}

\newcommand{\ie}{\textit{i.e.,}\xspace}

\usepackage[normalem]{ulem} % for \sout
\usepackage{xcolor}

\newcommand{\del}[1]{} % please delete
\begin{document}

%\title{Reinforcement Learning-Based Framework for Intelligent User Interface Adaptation"}

%LA OTRA OPCION::

%%%%%%%% 
% Reinforcement Learning-Based Framework for the Intelligent Adaptation of User Interfaces
%%%%%%%%%
\title{Reinforcement Learning-Based Framework for the Intelligent 
\\Adaptation of User Interfaces}
%Intelligent User Interface Adaptation: A Reinforcement Learning Framework


\author{Daniel Gaspar-Figueiredo}
\affiliation{%
  \institution{Universitat Politècnica de València \& ITI}
  \department{}
  \city{Valencia}
  \postcode{46022}
  \country{Spain}}
\email{dagasfi@epsa.upv.es}

\author{Marta Fernández-Diego}
\affiliation{%
  \institution{Universitat Politècnica de València}
  \city{Valencia}
  \country{Spain}}
\email{marferdi@omp.upv.es}

\author{Ruben Nuredini}
\affiliation{%
  \institution{Heilbronn University of Applied Sciences}
  \city{Heilbronn}
  \country{Germany}}
\email{ruben.nuredini@hs-heilbronn.de}

\author{Silvia Abrahão}
\affiliation{%
  \institution{Universitat Politècnica de València}
  \city{Valencia}
  \country{Spain}}
\email{sabrahao@dsic.upv.es}

\author{Emilio Insfrán}
\affiliation{%
  \institution{Universitat Politècnica de València}
  \city{Valencia}
  \country{Spain}}
\email{einsfran@dsic.upv.es}

\renewcommand{\shortauthors}{Daniel Gaspar-Figueiredo, et al.}

\begin{abstract}
  Adapting the user interface (UI) of software systems to meet the needs and preferences of users 
  % and to fit diverse contexts 
  is a complex task. The main challenge is to provide the appropriate adaptations at the appropriate time to offer value to end-users. Recent advances in Machine Learning (ML) techniques may provide
  effective means to support the adaptation process. 
  In this paper, we instantiate a reference framework for Intelligent User Interface Adaptation by using Reinforcement Learning (RL) as the ML component to adapt user interfaces 
    %In particular, a reference framework for Intelligent User Interface Adaptation has been proposed in a previous work. In this paper, we instantiate this reference framework by using Reinforcement Learning (RL) as the ML component to adapt user interfaces 
  %to various usage contexts 
  and ultimately improving the overall User Experience (UX). By using RL, the system is able to learn from past adaptations to improve the decision-making capabilities. Moreover, assessing the success of such adaptations remains a challenge.
  % for this type of systems. 
  To overcome this issue, we propose to use predictive Human-Computer Interaction (HCI) models to evaluate the outcome of each action (\ie adaptations) performed by the RL agent.
  In addition, we present an implementation of the instantiated framework, which is an extension of OpenAI Gym, that serves as a toolkit for developing and comparing RL algorithms. This Gym environment is highly configurable and extensible to other UI adaptation contexts. The evaluation results show that our RL-based framework 
  can successfully train RL agents able to learn how 
  to adapt UIs in a specific context to maximize the user engagement by using an HCI model as rewards predictor. 

\end{abstract}

\begin{CCSXML}
<ccs2012>
   <concept>
       <concept_id>10011007.10011074.10011075.10011077</concept_id>
       <concept_desc>Software and its engineering~Software design engineering</concept_desc>
       <concept_significance>500</concept_significance>
       </concept>
   <concept>
       <concept_id>10003120.10003123.10010860.10010858</concept_id>
       <concept_desc>Human-centered computing~User interface design</concept_desc>
       <concept_significance>300</concept_significance>
       </concept>
 </ccs2012>
\end{CCSXML}

\ccsdesc[500]{Software and its engineering~Software design engineering}
\ccsdesc[300]{Human-centered computing~User interface design}

\keywords{Adaptive User Interfaces, Reinforcement Learning, Human-Computer Interaction}

\received{24 February 2024}

\maketitle

\section{Introduction}

Adapting user interfaces (UIs) to the dynamic needs and preferences of users 
% and various contexts at 
taking into account the various contexts of use by suggesting changes at
the right time and place is a major challenge in software systems. The main goal is to provide timely and contextually relevant adaptations that significantly improve the User Experience (UX). 
Recent advances in Machine Learning (ML) have introduced promising ways to enhance the adaptation process~\cite{abrahaoModel:2021}. 
% Furthermore, ML is a very broad field so in this paper 
Thus, in this paper, 
we explore the use of 
% Reinforcement Learning as a component of ML for adapting IUs, 
RL as the ML component of the conceptual framework for
Intelligent User Interface Adaptation proposed in a previous work~\cite{abrahaoModel:2021}. 
%  providing the system with the ability to learn and optimise user interface adaptations based on previous adaptations. 
% \daniel{Maybe define here something more about RL as motivator?}
%The main motivation for using RL is to take advantage of the ability to learn how users interact with different interfaces and to be able to offer them changes or adaptations over time that improve the overall user experience. 
In this context, Reinforcement Learning (RL) techniques can be used as an instrument for step-wise adaptations of the UI based on user's interactions. This adaptation process will continuously align the interface with user's preferences improving the overall UX.
Additionally, the inherent "trial-and-error" nature of RL methods as well as the mechanism for penalizing mistakes and rewarding successes, provides better understanding of users' preferences. 
% In addition, such systems are able to learn from mistakes and successes, so that after each adaptation, the user can be better understood and eventually better adaptations can be provided to increase the UX.

Since RL is based on learning by success and failure, quantifying success in the context of UI adaptation is not a trivial task~\cite{sutton:1998RL}. 
% Successful adaptation can be considered to be that which improves user efficiency, or that which improves engagement, among others, but how to measure this is complex. 
Successful adaptation can be interpreted in different ways, for example, one that improves user efficiency, one that improves user engagement, or a combination of both. But how to assess these metrics and even how to combine them is far from straightforward.
Therefore, in this paper we propose the integration of predictive Human-Computer Interaction (HCI) models 
% as a measure of the success of adaptations.
to assist in this regard.
These models can provide an indicator of the success of adaptations by measuring certain aspects of user interaction and evaluating
% These models serve as tools to measure certain aspects of users and to evaluate 
the impact of adaptation actions on the overall UX. 
% More specifically, we use HCI models focused on predicting engagement as the starting point for this framework. 
As an initial approximation, our HCI model focus on predicting user engagement.

Finally, we present an implementation of the extended framework, based on OpenAI Gym. This implementation serves as a configurable toolkit that allows developers to create, on the one hand, the definition of the adaptive capabilities of their UI and the contextual information they can monitor and, on the other hand, to create and compare RL algorithms for the adaptation of the UI for the context they have defined. The configurable and extensible nature of the implementation ensures scalability for various scenarios.

Moreover, we evaluated this implementation by instantiating a specific case and running the evaluation in simulated environments. This experiment, though simulated, serves as a proof of concept, demonstrating the feasibility and adaptability of the proposed approach. Simulated data utilized in the experiment can be readily replaced with real-world or synthesized data, showcasing the flexibility and generalizability of the framework. The capability to adapt to different datasets while maintaining the same methodology highlights the versatility and potential real-world applicability of the extended framework.

This paper is organized as follows: In Section~\ref{sec:RW}, we review related works to contextualize our contribution. Section~\ref{sec:IUFramework} introduces the extended Intelligent User Interface Adaptation Framework, providing the conceptual basis. The implementation details of our RL agent for UI adaptation are presented in Section~\ref{sec:RLAgent}. Section~\ref{sec:Evaluation} offers an evaluation of the framework's performance. Finally, Section~\ref{sec:Conclusion} concludes the paper, summarizing key findings and suggesting potential avenues for future research.

\section{Related Work}
\label{sec:RW}
% \daniel{Estimated (1 page)}
% \silvia{faltaría citar el paper de Todi sobre adaptación de IU using RL discutiendo las difeencias con este paper y mencionndo nuestras contribuciones}
% \marta{Último párrafo de Related Works}

Adaptive systems and adaptive user interfaces (AUIs) have become crucial in modern software applications to tackle usability problems~\cite{Akiki:2014}. These systems can modify aspects of their structure, functionality, or interface to meet the varying and evolving requirements of individual users or user groups over time~\cite{Viano:2000}. However, providing appropriate adaptations, deciding when to present and display them, and ensuring they add value to end-users remains challenging for adaptive systems and AUIs. Different computational approaches to the problem of AUIs have been studied such as rule-based systems, heuristics, bandits, Bayesian optimisation, and supervised learning~\cite{Todi:2021MCTS}.

Research on AUI design has focused on the development of adaptation rules. These rules have traditionally been created with the help of UX experts' or system designers' knowledge~\cite{Hussain:2018}~\citet{Jean:2021a}. Moreover, most adaptive systems have relied on users directly stating their preferences. However, some facets of user preferences manifest through their behavioral patterns rather than through self-inspection. Recent approaches are increasingly considering implicit aspects of the user, such as their cognitive processing capabilities and the user’s physiological state~\cite{Chiossi:2022}. In the context of adaptive menus, systems still follow a heuristic approach where adaptations are selected based on manually-encoded rules that exploit data such as click frequency, visit duration or recency~\cite{Shrestha:2022}.

More recently, ML has enabled the automatic deduction of such adaptation rules from the users interaction data with the system. This automatic deduction process is performed using various ML techniques and algorithms \cite{Langley:1997}. Learning is recognized as a key capability for adaptive systems. For well-defined environments where the user state is highly predictive of adequate adaptation, the problem can be approached as a supervised learning problem.

In the more general situation where nor the user state is trivially known nor it is highly predictive of adequate adaptation, a RL approach is preferred. The multi-armed bandit problem is a classic RL problem that exemplifies the exploration–exploitation trade-off dilemma. Additionally, Bayesian optimization is used for problems with continuous parameters and an infinite number of potential options. In the context of AUI, such approaches offer a new paradigm for designing UIs in collaboration with Artificial Intelligence and user data \cite{Lomas:2016} \cite{Dudley:2019}. However, they have been proven successful in simple adaptation problems, such as recommendations and the calibration of interface parameters.

The special case of the RL problem in which the next state is not dependent on the action taken (\ie bandit problem) is not appropriate as regards learning policies for sequences of adaptations in which rewards are not immediately achievable. Problems that do require this kind of long-term planning should be solved with other RL algorithms. For example, Monte Carlo Tree Search (MCTS) has been proposed as a promising technique for the development of adaptive menu search interfaces \cite{Todi:2021MCTS}. In this context of menu searching in UI, \citet{Todi:2021MCTS} used predictive HCI models to predict rewards for each state during simulations. Since online simulations can be computationally expensive, a pretrained value network was used to directly obtain value estimates for unexplored states. Training data for this neural network was generated using the predictive HCI models. The authors showed that while the computation time increases drastically with simulations as search depth increases, it remains constant with the neural network approach without interfering much in the overall success rate (92.7\% with model-based simulation vs. 89.6\%).

Traditionally, AUIs have focused adaptations narrowly on specific UI elements rather than holistic changes across the entire system. According to the user's needs, preferences, and context, UI elements such as menus, buttons, bars, icons can be modified by adjusting font size, layout, color, contrast, theme. While this approach can efficiently resolve isolated usability problems, it disregards aspects of end-user's interaction that fall within the realm of UX. This inherent complexity of AUIs is imposing considerable challenges. Thus, UI adaptation should be considered as a multi-factorial problem in order to avoid improvements along some dimensions (e.g. user performance) at the expense of degradations along others (e.g. cognitive destabilisation) \cite{abrahaoModel:2021}. A related challenge is the resolution of conflicting UI adaptation alternatives. Once again, ML techniques could guide the multi-criteria decision-making process to identify the UI adaptation closest to the end-user’s goals and RL emerges as promising.

In fact, our purpose in this paper is not focused on a specific UI element and on a specific task, for example on menus to speed up searches as in~\cite{Todi:2021MCTS}, but on any adaptable UI element taking into account the needs and preferences of users. 
In our framework,
these adaptable UI elements can be easily added and removed as well as their different configurations
so, we provide a holistic view of the adaptation process of the UI. 
Unlike~\cite{Todi:2021MCTS} where the implementation of the RL environment is customized for the case of menu searches, our vision is to provide an environment that can be reusable for other purposes. Implementing it with OpenAI Gym enables this. Even our proposal will facilitate the integration of any RL algorithm to provide full modularity.

\section{Intelligent User Interface adaptation Framework}
\label{sec:IUFramework}

\begin{figure}
    \centering
    \includegraphics[width=0.48\textwidth]{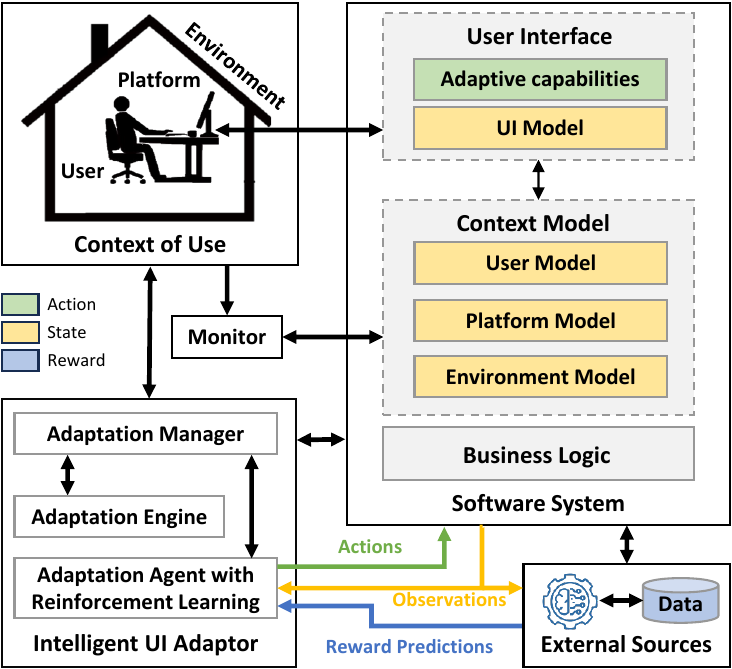}
    \caption{User Interface Adaptation framework using Reinforcement Learning. This is an extension from the original conceptual framework~\cite{abrahaoModel:2021}.
    } \label{fig:framework}
\end{figure}

% \begin{figure}
%     \centering
%     \includegraphics[width=0.46\textwidth]{framework_simplified.pdf}
%     \caption{User Interface Adaptation framework using Reinforcement Learning. This is an extension from the original conceptual framework~\cite{abrahaoModel:2021}.
%     } \label{fig:framework}
% \end{figure}

% This proposal is based on principles of Adaptive Systems.
%  % and predictive models of Human-Computer Interaction. 
% In particular, we explore the use of Reinforcement Learning, recently proposed to address user interface adaptation challenges. This technique is chosen for its ability to learn and infer knowledge without the need for initial training data.

In this paper we extend the conceptual framework proposed by
%~\citet{abrahaoModel:2021} 
Abrahão et al.~\cite{abrahaoModel:2021} 
by specifying what kind of ML techniques are going to be used and by implementing a first proof of concept. Since the conceptual framework is very complex and covers many aspects, we decided to implement only the mandatory elements in order to make a proof of concept. The Figure~\ref{fig:framework} shows the four parts of the generic framework (software system, context of use, intelligent UI adapter, and external sources) and how they are connected to each other.

The Software System includes a semantic core component that contains the \textit{Business Logic} functions specific to the application domain. The Software System also includes a \textit{User Interface}, which is responsible for presenting the functionality and data provided by the semantic core to the end-users.
This UI has some certain \textit{Adaptive Capabilities}, for example, redistributing elements or modifying the content displayed~\cite{jean:2004Graceful}. Moreover, this UI should be designed and defined as a \textit{UI Model}, for example, it could be defined using the Cameleon Reference Framework~\cite{calvary:2003CRF}. 

This software is intended to be run on a platform, in an environment and by a user.
In order to represent this, the \textit{Context of Use} represents the dynamic environment where end-users engage with the software system. 
It covers user-specific, platform-related and environment-related factors. This contextual information is then sensed by a \textit{Monitor} that senses and abstracts relevant data into the \textit{Context Model}. It can take into account many factors to get a better understanding on who, where and on what conditions the software is running. As an example, the \textit{User Model} could take into account factors such as emotional states, personality traits and other characteristics~\cite{heckmann2005gumo}. The Platform Model could include certain information about the device(s) where the Software is running on, and the Environment Model could include information regarding the surroundings where user and platform are located at.
This contextual understanding enables adaptive changes in the UI, responding to variations in the individual characteristics, platform specifications, or user's environment.

The conceptual framework proposed by~\citet{abrahaoModel:2021} defines multiple components in the \textit{Intelligent UI Adaptor}, but only few of them are mandatory to conduct the main adaptation process. In this paper, we focus only on the \textit{Adaptation Manager}, \textit{Adaptation Engine} and we extend the \textit{Adaptation Machine Learning} component to be an \textit{Adaptation Agent with Reinforcement Learning}:

\textbf{Adaptation Manager}: The Adaptation Manager is responsible for coordinating and managing the whole adaptation process. It also provides a UI that allows the end-user to interactively access and update adaptation parameters, review adaptation operations, and make decisions regarding the adaptation process. The Adaptation Manager also executes the adaptation logic contained in the \textit{Adaptation Engine} and coordinates the transition between the status before and after adaptation.
%through the Adaptation Transitioner. Additionally, it may request explanations for adaptation proposals or steps from the Adaptation Explainer and 
Additionally, it can integrate ML techniques to monitor and recommend adaptation operations based on other factors such as contextual changes and user interaction history.

\textbf{Adaptation Engine}: The Adaptation Engine contains the adaptation logic, which refers to the algorithm(s) used to perform the UI adaptation. This can include probabilistic-based models, rule-based approaches, case-based reasoning, logic-based methods, ontology-based techniques, evidence-based approaches, fuzzy logic, and more. The Adaptation Engine is responsible for processing the adaptation logic based on the current context and user requirements to generate adaptation proposals or decisions. 

%\textbf{Adaptation Agent with Reinforcement Learning}: This system monitors the adaptation process over time, learns from successful adaptations or user preferences, and recommends adaptation operations in the future based on this learning. It leverages machine learning techniques to improve the adaptation process and provide personalized adaptation recommendations to the end-user. 

\textbf{Adaptation Agent with Reinforcement Learning}: The Adaptation Agent with Reinforcement Learning operates within the framework of a Markov Decision Process (MDP) to continually monitor the adaptation process, assimilating feedback from successful adaptations and evolving user preferences over time.
The adaptation agent extracts essential decision-making information from the software system, namely state (observation) information from the \textit{Context Model} and the \textit{UI Model} and the actions from the \textit{UI Adaptive capabilities}. 
This agent operates on a reward-based mechanism, where \textit{external sources} contribute as reward predictors by taking advantage of other ML algorithms to process the data obtained by other sources. The historical data information is stored in the Data component so the RL agent can make more informed decisions. This user-centred reward mechanism guides the learning process, guiding the Adaptation Agent towards actions that positively impact UX.

% It takes advantage of machine learning techniques to improve the adaptation process and provide personalised adaptation recommendations to the end user.
% \marta{Aquí estamos combinando RL y ML. Si es a futuro, se puede explicitar y evita confusion: In the future, it could ... and provide more persolised}

%The External Sources are intended to supply the reward predictors for the Reinforcement Learning component. This external input enhances the learning process, making the system more adept at suggesting personalized adaptations. 

In the initial framework proposal, several properties were outlined, including the definition of the responsibility for adaptation and the underlying reasons for adaptation decisions. However, in our contribution we have chosen to centralise the responsibility for adaptation exclusively on the RL agent. This decision was made to simplify the proof of concept and to highlight the autonomous learning capabilities of the RL approach. It is worth mentioning that collaborative adaptation strategies will be further investigated, exploring synergies between users and the RL agent as future work. This collaborative approach is likely to involve shared decision-making, allowing users to contribute to the adaptation process, providing valuable insights and fostering a more interactive and user-centred adaptive system.

\section{Reinforcement Learning Agent to adapt User Interfaces}
\label{sec:RLAgent}

In this section we define the conceptualisation and implementation of our RL agent for adapting UIs. The conceptualisation is designed to cover a broad spectrum of states and actions to ensure flexibility and extensibility to multiple contexts. Subsequently, we present a practical example to demonstrate the instantiation of this conceptual framework, providing information on configuration and implementation. Finally, in the next section we will use this instantiation to evaluate the framework.

\subsection{Markov Decision Process Definition}

%% We have defined the UI adaptation problem as a stochastic sequential decision problem, where the adaptive system must plan a sequence of adaptations.

A Markov Decision Process (MDP) provides a mathematical framework to model decision-making processes in stochastic and dynamic environments,
% An MDP is a mathematical model for modeling decision-making processes in situations 
where outcomes are only partially influenced by the decisions that the agent takes. At its core, an MDP comprises states, actions, transition probabilities and rewards that guide the decision-making of an agent. These components collectively enable the agent to learn optimal strategies over time, making MDP a suitable approach for modeling and solving decision problems such as deciding which UI adaptations to perform and when.

\subsubsection{States}

The concept of states within the MDP framework encapsulate everything that may influence the decision-making process of an intelligent agent. In the case at hand, this entails the contextual information about the User, Platform and Environment and also the various UI design parameters.

In our framework, we represent the states using a \textit{MultiDiscrete} space to deal with different dimensions of information independently. 
%The agent receives this state representation as a vector that includes multiple dimensions, and each of these dimensions are assigned discrete values. 
The agent receives this state representation in the form of a vector that includes multiple dimensions, with each dimension assigned discrete values. 
By adopting this approach, the agent navigates and adapts within a well-organised state space, capturing the various factors that influence UI adaptations. This structured representation improves the comprehensibility and facilitates the configuration of the process.

One illustrative example of 
% the state changes 
changes in the state 
is when a user switches from a mobile to a desktop device (Platform change) while using an application. In this case, the state would show information changes such as screen size, device capabilities, operating system, and many others.
Another example is when a user transitions from a quiet office environment to a noisy coffee shop (Environment change). Here the state would reflect factors like ambient noise and ambient light levels 
% and potentially adapt 
to allow adaptations of 
the UI for better visibility or interaction. 
Moreover, if the user states some preferences about the UI configuration at a specific moment or reflects certain emotions while interacting with a software, which can both change over time (User change), these changes should also be updated in the state.

\subsubsection{Actions}

Actions cover a wide range of possibilities, from alterations in UI elements to more complex modifications in the overall UI design. These may include, but are not limited to, changes in colour schemes, repositioning of UI elements, resizing fonts, changing the content or even the navigation flow on the software~\cite{jean:2004Graceful}. These actions allows our system to respond intelligently to changes in user preferences, contextual variations and evolving environmental conditions.

\subsubsection{Transition probabilities}

It provides the probability of transitioning from one state to another state after performing adaptation. However, RL does not require explicit specification of the transition probabilities to solve MDP.

\subsubsection{Reward}
\label{sec:reward}

Defining a reward is usually a complex task, but even more for AUIs, due to both the nature of user interactions and the dynamic nature of interface adaptations. We recognise two key facets when defining the reward function: generality, which reflects general trends or preferences observed among users, and individual preferences, which take into account the unique choices of each user. To address these aspects in our reward, we have formulated the following reward function:

\begin{equation}\label{eq:reward}
    R = (1 - \sigma) \cdot G + \sigma \cdot I 
\end{equation}

Where, $G$ represents the general trends 
% that different types of users may have 
from different types of users
and $I$ represents the individual preferences of each user. 
% As an example, 
On the one hand, $G$ could represent the general tendency of users to prefer dark-themed interfaces~\cite{dakThemeCite}, based on extensive data analysis of user interactions. On the other hand, $I$ would encompass each user's unique preferences, such as preferring a larger font sizes or a specific colour scheme. Both $G$ and $I$ are normalized to a range between 0 and 1.
Then, the parameter $\sigma$, ranging from 0 to 1 too, allows us to adjust the balance of reward; for example, if $\sigma$ is closer to 0, generality is emphasised, while if $\sigma$ is close to 1, individual preferences are given more weight. Since every component is ranged from 0 to 1, the resulting reward is also ranged from 0 to 1. Thus, a reward closer to 1 means that the action (or adaptation) has been successful.

% that determines the balance between generality and individuality. If $\sigma = 1$, the reward is entirely determined by the generality component, and if $\sigma = 0$, the reward is solely determined by the individuality component.

\subsection{Implementation}

In our framework, the RL agent 
% acts as an engine for adaptation decisions, 
continuously monitors user interactions, learning from successful adaptations and recommends personalised adjustments. To implement the RL component, we opted for OpenAI Gym~\cite{brockman:2016openaigym}, a widely used toolkit designed to easily develop environments through a standardised interface and compare RL algorithms from the ones supported. 
% The choice of OpenAI Gym is due to its advantages, such as a standardised interface for defining environments, and support for several Reinforcement Learning algorithms. 
% The use of OpenAI Gym in our implementation provides a flexible basis that allows experimentation and benchmarking of our intelligent user interface adaptation framework based on Reinforcement Learning. 

Specifically, we created a customised OpenAI Gym environment for the AUI problem. This environment contains the representation of the states and actions and the design of the reward of AUIs, 
% providing an environment that represent the AUI problem 
all required 
for our RL agent.
To ensure compatibility with a variety of AUIs, our framework provides a configurable environment. This adaptability is facilitated through a configuration file, which allows developers to define the contextual information, (\ie User, Platform and Environment data) and the specific UI design information, in order to adjust the framework to their AUI. Moreover, developers can define the set of actions that the UI can take in response to dynamic contextual changes.
Furthermore, the configurability covers also the design of a reward model, enabling developers to shape the system's learning process by specifying the rewards associated with the adaptation outcomes.
% This is, if all the key components about \textit{States}, \textit{Actions} and, \textit{Reward}, from Figure~\ref{fig:framework} are defined, then everything will be setted up and created to start with the Learning of the RL Agent.

% The configuration file should be provided in a specific format and then the environment to train RL agents will be automatically set up. Once the environment is ready, the training of agents using any of the compatible RL algorithms can start. 

% In our framework, we enable connectivity 
On the other hand, our framework provides connectivity 
with AUIs through API calls. 
% The assumption is that if a software system incorporates adaptive capabilities, these functionalities should be easily accessible 
When a software system incorporates adaptive functionalities, these should be easily accessible
through well-defined API calls
%. This connectivity allows 
allowing
the intelligent adapter, which can operate remotely, to activate or use the adaptive functions of the AUI. 
% We have designed this feature with flexibility in mind, allowing developers to 
Developers can specify API calls directly within the configuration file to articulate the exact functionalities and actions that the intelligent agent can call on the AUI. This approach ensures that the adaptive process is not limited only to a specific environment, but can dynamically interact with and influence external AUIs based on the specified API calls. 
The complete implementation is available at \url{https://github.com/RESQUELAB/RL-UIAdaptation}

\subsection{Setup}
\label{sec:setup}

Since our MDP definition is general to the AUI problem, it can be instantiated with different specific states, actions and reward sources. We define in this section an example of MDP specification that will be used to evaluate our framework.    

% Since our implementation is configurable and extensible to multiple types of states, actions and reward sources, we show here an example that will be used to evaluate the framework. 
% The constraint is that we have to provide \textit{i)} the contextual information that will be taken into acount for the adaptation; \textit{ii)} the adaptive capabilities of the system and; \textit{iii)} the sources of the reward in both $G$ and $I$ aspects. 
% In this section we described how the MDP, but it does not have any specific values for the Contextual information and the UI design. In order to shed some light into these concepts, we proceed with a more specific representation of states

\subsubsection{States}
\label{sec:setupStates}
The state includes both, the UI design information and the contextual (User, Platform and, Environment) information. In our first prototype, simple but able to demonstrate the functionality of the framework, only the UI design features and the User preferences will be considered. Future iterations of the framework will include the variability over the Platform and Environment.
% As our initial goal is to start with simple cases and iteratively improve the framework, we are going to consider only the UI design characteristics and the User preferences as information that can vary and should be taken into account in order to adapt the UI. We keep the variability over the Platform and Environment aspects for future improvements. % Future iterations of the framework may include more variables to capture additional dimensions of the user experience.

With regard to the UI design information that we considered for this evaluation, we focused on four variables: \textit{Layout}, \textit{Theme}, \textit{Font size} and \textit{Information display}. These variables represent the structure, appearance and contents of the interface.
The layout variable can take the values \textit{list} and a range of \textit{grid} layouts, from \textit{2-column grid} to \textit{5-column grid} changing how the elements distribute over the interface, the \textit{theme} can take the values \textit{light} and \textit{dark}, the \textit{font size} can be \textit{small}, \textit{default} or \textit{big} and, \textit{information display} can be \textit{show}, \textit{partial}, and \textit{hide}, which will show, show partially or hide some information to the users.
On the user side, we took into account changes in \textit{preferences}. The user's preferences can take the values of the design of the UIs.
Given the four UI design variables (layout, theme, font size, and information display), each with multiple possible values, the total number of possible combinations can be determined by multiplying the number of options for each variable, resulting in $5\cdot2\cdot3\cdot3=90$ combinations. Moreover, we must consider the user preferences with the same possible values as for the UI design. Consequently, this leads to $(5\cdot2\cdot3\cdot3)^{2} = 8100$ possible states in this context.

% Given the four UI design variables (layout, theme, font size, and information display), each with multiple possible values, and considering the user preferences with the same possible values as for the UI design, the total number of possible combinations can be determined by multiplying the number of options for each variable. For instance, $(5\cdot2\cdot3\cdot3)^{2} = 8100$ possible states.

Future evaluations may include additional variables in both UI design and contextual representations to comprehensively assess the adaptability of the framework across various setups.

\subsubsection{Actions}

The actions considered in this MDP instantiation correspond to changes in the UI design variables. Specifically, we consider actions to distribute elements in a \textit{0) list}, or \textit{1) grid} layout with 2-columns, \textit{2) grid} layout with 3-columns, \textit{3) grid} layout with 4-columns and \textit{4) grid} layout with 5-columns; to activate the \textit{5) light} or \textit{6) dark} themes, to change how the characters are displayed on the UI with a \textit{7) small}, \textit{8) default} or \textit{9) big} font sizes, and change the amount of information displayed on the screen by \textit{10) showing} \textit{11) showing partially} and \textit{12) hiding} the information. Additionally, we introduce a \textit{13) No operate} action, allowing the agent not to make any change. This action acknowledges that, in some situations, maintaining the current UI configuration might be the optimal decision for the agent.
To simplify the representation and for compatibility with OpenAI Gym interface, we map these actions to numerical values ranging from 0 to 13, resulting in a total of 14 actions that the agent can take.

\subsubsection{Reward}

The reward proposed in the Section~\ref{sec:reward} is defined by Equation~\ref{eq:reward}. This Equation has two main components, the Generality ($G$) and Individuality ($I$). 
However, this equation does not specify the source or the way in which these components are derived. 
In this MDP instantiation, we rely on a predictive HCI model to obtain the $G$ component.
Then, for the $I$ component, we calculate the alignment between the user preferences and the actual UI design configuration. This measures how similar the preferences are compared to the UI design.

On the one hand, with regard to the predictive HCI model, an experiment was conducted involving 25 master students to record their interactions with various UI configurations. 
    This experiment consisted in recording every interaction the participants did while using an e-commerce catalogue. We asked the participants to proceed with the purchase of some products using different configurations of UIs. Specifically, we registered interactions with the UI combinations of layout and theme (\ie layout in grid or list and theme in dark or light). 
% and we recorded the number of clicks, scrolls, and events in order to predict their engagement\cite{LEHMANN:2012, barbaro:2020, carlton:2021}.
    This experiment provided a data set of how many clicks, scrolls and events that the participants did for each of the configurations. This dataset was then used to calculate the engagement levels~\cite{LEHMANN:2012, barbaro:2020, carlton:2021} associated to the various UI configurations that the participants were using.

However, training a high-quality HCI model requires a substantial number of examples to ensure good performance on unseen, future data. The processes involved, such as recruiting participants, data collection, and labeling, are typically time-consuming and expensive. In response to these challenges, we applied data augmentation techniques to enhance the dataset obtained from the experiment. This approach allowed us to artificially increase the amount of interaction data, addressing the limitations posed by a small dataset. Specifically, we employed SMOTE~\cite{smote:2002}, a data augmentation technique, to generate additional instances, resulting in a more comprehensive dataset for training the predictive HCI model. In the end, we obtained 310 samples of interaction data.
%
%
    %Since the amount of data may be small, we decided to apply SMOTE, a data augmentation technique, which allowed us to obtain 340 interaction data.
%
Then we trained a 
    \textit{Random Forest Regressor} model capable of predicting a user's engagement level based on the UI configuration. 
    We evaluated the accuracy of the model and we obtained a mean squared error of $0.055$.

On the other hand, to calculate the alignment of the $I$ component, we created a \textit{simulated user} which has a set of predefined preferences. The alignment measure consists of comparing the preferences of the user with the actual configuration of the UI design. This distance assesses the degree of overlapping or agreement between the user's predefined preferences and the actual state of the UI. A higher alignment score implies a closer alignment, indicating that the UI design is more consistent with the preferences of the user.

\subsubsection{Algorithm}

    In our prototype, we selected Q-Learning from OpenAI Gym's suite of RL algorithms due to its simplicity, effectiveness with discrete actions, and widespread use as a baseline algorithm.
    Q-Learning is an off-policy algorithm which aims to determine the optimal next action based on the current state to maximize the accumulated rewards. The Q-value, denoted as $Q(s, a)$, represents the estimate of the cumulative future rewards  anticipated by the agent in state $s$ and taking action $a$. The Q-Learning update equation (Equation~\ref{eq:QLearning}) refines this estimate over time:

    \begin{equation}
    Q(s,a) = (1 - \alpha) \cdot Q(S_{t},A_{t}) + \alpha \cdot (R_{t+1} + \gamma \cdot \max_{a}Q(s_{t+1}, a))
    \label{eq:QLearning}
    \end{equation}
    
    Here, $Q(S_{t},A_{t})$ represents the current estimate of the Q-value, where $\alpha$ is the Learning Rate. The update term, $\alpha \cdot (R_{t+1} + \gamma \cdot \max_{a}Q(s_{t+1}, a))$, incorporates the Learning Rate, immediate reward ($R_{t+1}$), Discount Factor ($\gamma$) for future rewards, and the maximum Q-value for the next state $s_{t+1}$. This equation guides how the agent refines its understanding of Q-values in order to make more informed decisions over time.

    In this context, \textit{episodes} represent complete iterations during training, with each \textit{step} corresponding to an individual action or decision made by the agent (\ie adaptations). Equation~\ref{eq:QLearning} guides the agent in refining its understanding of Q-values over steps within an episode.
    
    Additionally, the Q-Learning algorithm starts with no knowledge and balances exploration and exploitation with the \textit{Exploration Factor} ($\epsilon$). As episodes progress, $\epsilon$ is gradually reduced, shifting the agent from exploration to exploitation based on accumulated knowledge.
    
\section{Evaluation}
\label{sec:Evaluation}

According to the Goal-Question-Metric (GQM) template for goal definition~\cite{Basili:1994:GQM}, the goal of this study is to
\textit{analyze} the 
Reinforcement Learning agent
% intelligent user interface adaptation framework based on Reinforcement Learning
\textit{with the purpose to} evaluate its effectiveness 
\textit{with respect to} the ability to learn to adapt the UI  
\textit{from the point-of-view of} researchers when training RL agents
\textit{in the context of} simulated environments, with UI design and User preferences variability and reward obtained through an HCI model trained with historical interaction data obtained from a previous experiment.

This evaluation is conducted through simulations to assess the framework's performance in a controlled environment. The simulations are executed on a computer with the following specifications: Intel Core i9-13900KF, 64 GB DDR5, NVIDIA GeForce RTX 4090. 
The simulations are set up with the specifications defined in Section~\ref{sec:setup}.
These simulations aim to provide insights into the framework's capabilities and limitations, for further development and improvement. 
It's important to note that while simulations serve as a valuable starting point, future work should extend this evaluation process to include human interactions.

\subsection{Configuration Parameters of the RL Algorithm}

The Equation~\ref{eq:QLearning} presents multiple parameters that influence on how the agent will perform its learning. Our goal is to ensure optimal balance between exploration and exploitation in the learning process. The choice of parameter values is based on published experiences applying Q-Learning to tasks with comparable complexity. Furthermore, expert recommendations played a pivotal role in refining and finalizing these values:

\begin{itemize}
    \item Number of episodes (60000): The total number of iterations or episodes the Q-Learning algorithm will go through during training. 
    
    \item Learning Rate ($\alpha = 0.90$): A relatively high Learning Rate allows the agent to adapt quickly to new information. A higher Learning Rate ensures that the agent incorporates recent experiences effectively. 
    % A value of 1 has not been used because it is not wanted to completely discard prior learning.
    
    \item Discount Factor ($\gamma = 0.90$): The Discount Factor determines the weight given to future rewards. A value of 0.90 prioritizes future rewards while still considering short-term consequences.

    \item Exploration Factor, $\epsilon$: Has an initial value of 1 with linear decay in the first half of the training process (30000 episodes) until its value reaches $0.1$. This decay heuristics promotes more exploration in the early stages and exploitation in the later stages of training. The minimum value of 0.1 allows the agent to still explore sometimes while exploiting.
    
    % \item
    %Generality \textit{vs. } Individuality trade-off ($\sigma$): In order to explore the trade-off between $G$ and $I$ in Equation~\ref{eq:reward}, we set the $\sigma$ to multiple the values of [$0$, $0.25$, $0.50$, $0.75$, $1$], providing insights into how the system responds to different degrees of personalization.

% Finally, since we want to explore the possibilities of prioritizing individuality over generality and viceversa, we set the $\sigma$ to multiple the values of [$0$, $0.25$, $0.50$, $0.75$, $1$]. 
% This way, we could evaluate if adjusting the parameter $\sigma$ influences the 
% % balance 
% trade-off 
% between generality and individuality in the reward function, providing insights into how the system responds to different degrees of personalization.

\end{itemize}

Finally, we aim to explore the trade-off between the $G$ and $I$ in Equation~\ref{eq:reward}, where we set $\sigma$ to various values [$0$, $0.25$, $0.50$, $0.75$, $1$], offering insights into the system's response to different levels of personalisation.

\subsection{Evaluation Metrics}

% In order to evaluate the ability to learn and the efficiency of the agent used for this framework implementation, the following metrics, typically used to evaluate RL agents, have been selected: 

In order to evaluate the ability to learn and the efficiency of the agent used for this framework implementation, we have selected the following metrics, which are commonly used to assess RL agents~\cite{mnih2013playing, Mnih:2015}:

\begin{itemize}
    \item 
        \textit{Number of steps}: 
            It reflects the total count of adaptations taken by the RL agent to achieve a fully adapted UI. The purpose of tracking the number of steps is to assess the efficiency and effectiveness of the learning process. Ideally, a well-performing agent should learn to adapt the UI minimizing the number of steps required for a complete adaptation. A higher number of steps may indicate suboptimal learning.
        
    \item 
        \textit{Episode Score}:
        The episode score represents the cumulative reward obtained throughout a single episode. The episode score reflects the  effectiveness of the agent's actions during that episode. Higher episode scores indicate better performance, as they suggest that the agent has made decisions leading to more favorable outcomes.
\end{itemize}
            
    %An improvement of the score over time would mean that the agent is learning correctly.

To address the inherent noise and fluctuations in the learning process visualization for both the \textit{Number of steps} and \textit{Episode Score}, we incorporated an anti-jittering~\cite{pong:2022metaRL} parameter. This parameter involves recording average values at regular intervals. In our case, we calculate the moving average over a window of size 150 episodes, providing a stable and representative trend for assessing the performance of the RL agent's learning. This approach ensures that both metrics are presented in a smoothed manner, making it easier to interpret the overall learning progress.
                
    % Additional to the algorithm parameters, we considered an anti-jittering parameter which addresses the noise or fluctuations in the visualisation of the learning process~\cite{pong:2022metaRL}. By recording average values every 150 episodes, the reported results become more stable and representative of the overall learning trend.
    
    % Given the fluctuating nature of this score, we utilize a moving average of its last values within a window of size governed by the anti-jittering parameter. In our case, the visualization of the episode score is the average of the last 150 episode scores. Plotting these averages provides smoothed representation of the overall performance trend helping to assess the performance of the RL agent's learning. 

In addition to the commonly used metrics in evaluating RL agents, we propose an additional metric:

\begin{itemize}
    \item 
        \textit{Alignment Score}:
        The alignment score measures how well the RL agent's adaptive decisions align with user preferences, providing insights into personalized UI adaptation effectiveness. It's normalized from $0$ to $1$, calculated as:
        % The alignment score aims to capture how well the adaptive decisions made by the RL agent align with the preferences of the users. The alignment score provides valuable insights into the effectiveness of the framework in adapting UIs in a personalized manner. It's value is normalized from $0$ to $1$ and its calculated following this equation:

            \begin{equation*}
                Alignment = \frac{Max Alignment - Mismatched Attributes}{Max Alignment}
            \end{equation*}
            
            Where $Max Alignment$ is the total number of attributes in the user preferences, and \\$Mismatched Attributes$ is the count of attributes where the user preferences differ from the UI design.    
\end{itemize}

\subsection{Results}

In this section, 
we distinguish between evaluation during learning which is computed over the course of training and evaluation after learning, which is evaluated on a fixed policy after the agent has been trained. During the learning process (\ie training), evaluation provides insights into how the RL agent evolves over time, demonstrating its ability to adapt and optimize UI configurations. Subsequently, 
once the agent is trained, the evaluation process (after learning) assesses the agent's performance on simulated users with specific preferences.

\subsubsection{Learning Process}

In the learning process, we ran 60000 episodes for each of the different values for $\sigma$ to train the RL agent. We included an extra reward of 1 if the agent finished the whole adaptation process in 4 or less steps. Since there are four types of actions, we presumed that optimal adaptation can be achieved in four steps.

Each episode started with the creation of a new random initial state. Consequently, a new simulated context was generated in each episode, with a random user with specific preferences and new UI design features selected from the available options. The agent's goal was to adapt the UI to maximize the reward, which depended on the $\sigma$ value. 
%
% A low $\sigma$ (close to 0) emphasized the outcome predicted by the HCI model, while a high $\sigma$ (close to 1) prioritized the alignment between user preferences and UI design.

\begin{figure}
    \centering
    \includegraphics[width=0.45\textwidth]{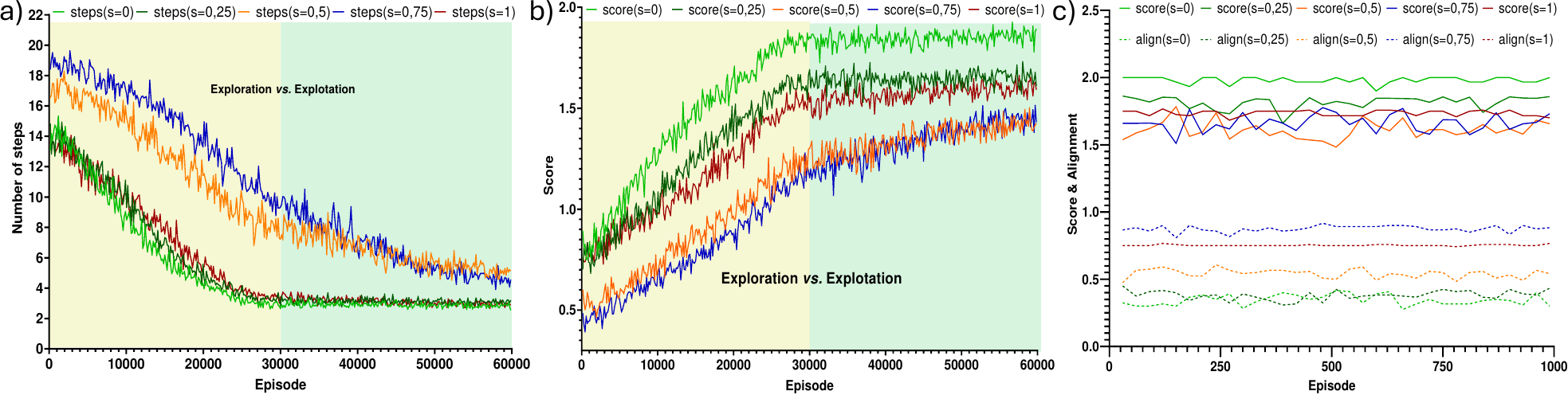}
    \caption{
        RL agent learning process. a) The number of steps needed to finish an episode decreases over time;  b) The score increases over time and the agent converges to an optimal solution; c) RL agent evaluation process. 
    } \label{fig:results}
\end{figure}

As illustrated in Figure~\ref{fig:results}~\textit{a)} and \textit{b)}, the agent demonstrates convergence across various values of $\sigma$, transitioning from an exploration phase (depicted in yellow) to an exploitation phase (depicted in green) after the initial 30,000 episodes.
% In particular, the exploration phase allows the agent to discover effective adaptive strategies and, subsequently, in the exploitation phase, the agent refines its actions based on the acquired knowledge.
However, it's worth noting that for $\sigma = 0.5$ and $\sigma = 0.75$, complete convergence may not have been achieved, suggesting ongoing exploration and refinement of adaptive strategies even in later episodes.

The analysis of the \textit{number of steps}, as shown in Figure~\ref{fig:results}~\textit{a)}, reveals that the agent efficiently completes episodes in 3 to 4 steps across most scenarios, indicating an adaptation strategy focused on essential changes. However, for $\sigma = 0.5$ and $\sigma = 0.75$, the agent required an additional step, taking 4 to 5 steps to conclude the episodes. This deviation suggests that when there is a balance between generality and individuality leaning towards individuality may lead to increased complexity in the adaptation process.

% The analysis of the \textit{number of steps}, as shown in Figure~\ref{fig:results} \textit{a)}, the number of steps converge to an optimal number in almost all scenarios. This means that the agent efficiently completes
% the episodes in 3 to 4 steps, demonstrating an adaptation strategy focused only on essential changes. However, for $\sigma = 0.5$, there is a big deviation where the agent needed around 12 steps to conclude the episodes. This anomaly suggests that when there is a balance between generality and individuality, the agent may have difficulty distinguishing optimal sequences from adaptations.

Meanwhile, examination of the \textit{score} obtained over time, as depicted in Figure~\ref{fig:results}~\textit{b)}, offers insights into the agent's learning progress and performance. Across most scenarios, a steady growth trajectory is observed, indicating that the agent improves its performance over time. However, notable differences in the score are evident depending on the $\sigma$ value. For $\sigma = 0$ and $\sigma = 0.25$, where the reward is predominantly influenced by generality, higher and more stable rewards are achieved. This observation suggests that prioritizing generality in the reward function leads to more consistent performance and effective adaptation strategies. Conversely, for intermediate values of $\sigma$, such as $\sigma = 0.5$ or $\sigma = 0.75$, where individuality is prioritized, the reward tends to be lower and less stable. 
However, in the case of $\sigma = 1$, the agent also obtains a high and stable reward. 
This phenomenon highlights the challenge of balancing generality and individuality in the adaptation process, as reflected in the fluctuating nature of the score. Moreover, the influence of $\sigma$ on the reward function underscores the importance of carefully selecting and tuning this parameter to optimize the agent's performance in adapting user interfaces.

% Meanwhile, examination of the score obtained over time, as shown in Figure~\ref{fig:results}~b), reveals a steady growth trajectory that converges for  $\sigma$ values. However, the impact of $\sigma$ on the reward function is notable. 
% Notably, the divergence in reward values for the $\sigma = 0.5$ scenario may indicate a unique challenge when balancing generality and individuality.
% Nevertheless, values close to 0 and 1, where the reward is weighted to generality and individuality, respectively, produce a higher reward. 
% % 
% % 
% % 
% This trade-off in $\sigma$ values suggests that finding an optimal adaptation strategy becomes more complex in scenarios where there is a balance between both aspects at the same time.

% \begin{figure}
%     \centering
%     \includegraphics[width=0.45\textwidth]{Learning_steps.pdf}
%     \caption{
%         Learning process of the RL agent. The number of steps needed to finish an episode decreases over time.
%     } \label{fig:learning_steps}
% \end{figure}

% \begin{figure}
%     \centering
%     \includegraphics[width=0.45\textwidth]{Learning_graph.pdf}
%     \caption{
%         Learning process of the RL agent. The score increases over time and the agent converges into an optimal solution.
%     } \label{fig:learning}
% \end{figure}

\subsubsection{Evaluation Process}

We used the previously trained agent in a series of 1000 episodes to evaluate its performance. Similar to the learning process, each episode in this phase involved creating simulated users with random preferences and starting the adaptation process from a randomly generated initial UI configuration.

Figure~\ref{fig:results} c) illustrates the agent's performance in terms of score and alignment across the 1000 episodes for each value of $\sigma$. 
We observe that \textit{alignment} increases as $\sigma$ emphasizes individuality, with the highest alignment values observed for $\sigma$ values of $1$ and $0.75$. Conversely, the lowest alignment values are associated with $\sigma$ values of $0$ and $0.25$, while $\sigma = 0.5$ exhibits intermediate alignment, as anticipated.
Regarding the \textit{score},
notably, $\sigma = 0.5$ and $\sigma = 0.75$ exhibit more fluctuations compared to other values, suggesting potential for further refinement and convergence through parameter adjustments in future simulations.
%
%
%
% we
We find that $\sigma$ values of $0$ and $0.25$ yield the highest scores. This indicates that emphasizing generality in the reward function leads to higher cumulative rewards. However, it is noteworthy that all $\sigma$ values achieve good scores, with values consistently exceeding 1.5 across all scenarios. Despite some variations, these results demonstrate the effectiveness of the agent in adapting user interfaces across different $\sigma$ values.

% We illustrate the agent's performance in terms of score and alignment over the 1000 episodes in Figure~\ref{fig:results} c). As expected, when $\sigma=1$, the agent prioritized individual user preferences, resulting in higher alignment scores. In contrast, for $\sigma=0$, the agent emphasized generality using the HCI model, leading to lower alignment. However, the noteworthy observation arises when $\sigma=0.5$, which represents a balanced consideration of both individual and general preferences. In this case, the agent achieved alignment values close to 1, along with a good overall score. This finding suggests that achieving a balance between individual and general factors is advantageous, as it yields high alignment while incorporating general knowledge from the HCI model. 

% This evaluation suggests that the agent was able to learn and adapt to the different contexts. 

% \begin{figure}
%     \centering
%     \includegraphics[width=0.45\textwidth]{Evaluation.pdf}
%     \caption{Evaluation process of the RL agent. Score and alignment (align) outcomes for different $\sigma$ values. Note that when $\sigma = 1$, the $I$ 
%     % component
%     in the reward function has the same value as the \textit{alignment}.} \label{fig:evaluation}
% \end{figure}

\subsection{Threats to Validity}

In this section, we discuss the threats to validity of our research, following the guidelines from~\cite{Wohlin2012Experimentation}:

\textbf{Internal Validity}: One potential threat is that states or transitions remain unexplored during the agent's learning process.
% We chose a small state space for the proof of concept to facilitate experimentation, ensuring the Q-table covered all possible transitions. 
We deliberately constrained the size of the state space. This deliberate choice to limit the complexity of the state space facilitated more efficient experimentation and ensured thorough exploration of all possible transitions within the Q-table.
Another internal threat concerns reward definition, which may not perfectly quantify adaptation success. However, we designed a generic reward adaptable to each context.
% , testing various parameters in a specific scenario with expected behavior. 
The use of a predictive HCI model for the generalistic ($G$) part of the reward introduces an internal challenge. The accuracy of the HCI model depends on the dataset quality; our model, trained with data augmented using the SMOTE method from 25 participants, resulting in a dataset of 310 samples, achieved a mean squared error of $0,055$. Despite the limited participant number and data quantity may impact HCI model generalization.

\textbf{External Validity}:
An external threat comes from the simplicity of our contextual representation, focusing only on user changes and neglecting platform and environmental variations. Generalizing our results to more complex contexts may be limited. To enhance external validity, future experiments should include broader contextual factors, ensuring a more comprehensive evaluation of our RL-based UI adaptation framework.
Additionally, relying on the HCI model to predict user engagement introduces another external threat. While our current experiments use a simplified context representation, extending the model to predict additional aspects like user satisfaction or cognitive load could enrich contextual information. This improvement would offer a more complete representation of system state variability, contributing to a more realistic assessment.
Furthermore, our choice of the Q-Learning algorithm introduces a potential external threat. While Q-Learning is well-established, exploring alternative RL algorithms, such as Proximal Policy Optimization (PPO)~\cite{schulman:2017ppo} or Monte Carlo Tree Search (MCTS), in future research may provide a broader understanding of the framework's performance across different algorithmic approaches.

\section{Conclusions and Future Work}
\label{sec:Conclusion}

% \daniel{Talk about the "ALIGNMENT" problem in AI... which is the idea of aligning the goals and behavior of artificial intelligence systems with human values and objectives. In our case, how can we define an Agent that effectively does what we want to achieve? --> Maybe link this to the HCI models + User preferences? }
% \marta{Igual mejor enlazar con este concepto en la conclusión, no aquí.}

This paper presents an instantiation of a reference framework for Intelligent User Interface Adaptation utilizing Reinforcement Learning as the machine learning component. The framework aims to adapt user interfaces, ultimately enhancing the overall User Experience. 
By formulating the problem using MDP, we provide a comprehensive representation of the adaptation process. Our framework offers high configurability and expandability, facilitating customization for specific use cases and evolving requirements. We introduce a holistic reward function capturing both general trends and individual user preferences, alongside a flexible $\sigma$ parameter for fine-tuning adaptation. Simulated evaluations offer valuable insights into framework performance, but real-world validation is indeed needed. Future work includes exploring alternative reward mechanisms, such as human feedback~\cite{figueiredo2023comparative} and algorithms, such as PPO and MCTS, enhancing state coverage, and improving predictive HCI models for better context representation.

\begin{acks}

This work is supported by the AKILA project (CIAICO/2021/303)
funded by the GVA and UCI-Adapt project (PID2022-140106NB-I00) funded by the AEI. D. Gaspar-Figueiredo is funded by the GVA (ACIF/2021/172), which is cofunded by the European Union through the ESF.

\end{acks}

\bibliographystyle{ACM-Reference-Format}
\bibliography{main}

% \appendix
\end{document}